\documentclass[aps,pra,reprint,onecolumn,notitlepage,eqsecnum,floatfix,showkeys,nofootinbib]{revtex4-2}
\usepackage{amsmath}
\usepackage{amsbsy}
\usepackage{xcolor}
\usepackage{graphicx}
\usepackage{hyperref}
\usepackage{multirow}
\usepackage{linearb}
\usepackage{relsize}
\usepackage{calrsfs}
\DeclareMathAlphabet\mathbfcal{OMS}{cmsy}{b}{n}

\newcommand{\bq}{\begin{eqnarray}}
\newcommand{\eq}{\end{eqnarray}}
\newcommand{\bqn}{\begin{eqnarray*}}
\newcommand{\eqn}{\end{eqnarray*}}
\newcommand{\bqs}{\begin{subequations}}
\newcommand{\eqs}{\end{subequations}}
\newcommand{\bw}{\begin{widetext}}
\newcommand{\ew}{\end{widetext}}

\newcommand{\kk}{{\boldsymbol k}}

\newcommand{\rr}{{\boldsymbol r}}

\newcommand{\nnabla}{{\boldsymbol\nabla}}

\newcommand{\calt}{{\cal T}}

\begin{document}
%%%%%%%%%%%%%%%%%%%%%%%%%%%%%%%%%%%%%%%%%%%%%%%%%%%%%%%%%%%%%%%%%%%%%%%%%%%%%%
%%%%%%%%%%%%%%%%%%%%%%%%%%%%%%%%%%%%%%%%%%%%%%%%%%%%%%%%%%%%%%%%%%%%%%%%%%%%%%
%%%%%%%%%%%%%%%%%%%%%%%%%%%%%%%%%%%%%%%%%%%%%%%%%%%%%%%%%%%%%%%%%%%%%%%%%%%%%%
\title{Temperature of the Vacuum}

\author{Riccardo Fantoni}
\email{riccardo.fantoni@scuola.istruzione.it}
\affiliation{Universit\`a di Trieste, Dipartimento di Fisica, strada
  Costiera 11, 34151 Grignano (Trieste), Italy}

\date{\today}

\begin{abstract}
In a recent trilogy we proposed a Statistical Theory of General Relativity 
spacetime. Here we apply our new theory to determine the (energy) ``density''
and (virial) ``temperature'' dependence of the structure of the spacetime 
quantum vacuum working on the simple case of a real massless scalar field
in a local Lorentz frame.
\end{abstract}

%\pacs{...}

\maketitle
%\tableofcontents
%%%%%%%%%%%%%%%%%%%%%%%%%%%%%%%%%%%%%%%%%%%%%%%%%%%%%%%%%%%%%%%%%%%%%%%%%%%%%%
\section{Introduction}
%%%%%%%%%%%%%%%%%%%%%%%%%%%%%%%%%%%%%%%%%%%%%%%%%%%%%%%%%%%%%%%%%%%%%%%%%%%%%%
\label{sec:intro}

In a recent trilogy \cite{Fantoni24f,Fantoni25a,Fantoni25g} we  
proposed a Statistical Theory of Gravity. This allowed us to determine a
``virial temperature'' of the spacetime metric tensor field. Albeit still under
refinement, the theory is already able to offer some measurable predictions.
In fact, as we will see in this work, it influences the energy density 
{\sl structure} of the spacetime vacuum. Unfortunately, with the current equipment
we are unable to directly measure these structure variations due to the 
temperature. But we can hope in some indirect observations of the consequences
on the Hubble rate of expansion of the Universe, the parameter $H=\dot{a}/a$, 
where $a>0$ is the scale factor which enters the spatial components of the 
cosmological metric tensor field and is proportional to the average separation 
between objects, such as galaxies, and the dot denotes a derivative with 
respect to the cosmological time.
As Edwin Hubble discovered in 1929 the parameter $H$ is a measurable 
quantity. For example the current Hubble parameter, the {\sl Hubble constant}, 
is estimated to be $H_0\approx 7\%/$Gyr. Hubble constant is made of two 
contributions: a gravitational one and one due to the cosmological constant.
Wheeler’s spacetime foam \cite{Wheeler1957,Gravitation} suggests that a 
foamy structure leads to the cosmological constant we see today.
Cosmological models for the metric tensor field began with the one of 
Friedmann-Lema\^itre-Robertson-Walker and were refined in various ways 
\cite{Wang2017} in order to take care of the inhomogeneity and anisotropy of 
spacetime predicted by a quantum vacuum. These calculations have macroscopic
consequences at the level of the description of the Universe evolution. 
In the sense that $a(t)$ can have various different functional forms: 
respect to the current situation it can grow linearly, it can grow 
exponentially or with other laws, or it can even bend downwards with some 
law resulting in a big crunch. The Universe should be seen as embedded in
a higher dimensional space, as the ``surface'' of a bubble nucleated at a 
certain point of our spacetime. The bubble can either grow or shrink. 

Through the Universe exploration we can hope to be able to at least 
have some indirect insight on the effect of temperature on the spacetime 
vacuum. One crucial step in our formulation is assuming that the vacuum 
energy density and its temperature are constant in cosmological time and uniform 
in cosmological space. In this respect the energy density of the spacetime vacuum
$\rho_{\rm vacuum}$ is conceptually different from the energy density of
matter $\rho_{\rm matter}$ in the Universe: while the matter mass has to 
be considered a constant during the Universe evolution so that 
$\rho_{\rm matter}\propto a(t)^{-3}$ the vacuum is created or destroyed
during the Universe expansion or contraction with $\rho_{\rm vacuum}$
kept constant. In other words the spacetime vacuum behaves like a fluid
occupying a larger or smaller volume but keeping its density 
constant. So it behaves like a cosmological constant in the Universe
evolution. We may consider this fluid as the source of dark energy.

Our virial temperature is conceptually different from the Davies-Unruh 
\cite{Davies1975,Unruh1976} local temperature. The latter is in fact defined as 
$\calt_{DU}=\hbar g/2\pi c k_B\approx 4.06\times 10^{-21}$ Ks$^2$m$^{-1}\times ~g$ 
where $g$ is the proper uniform acceleration of a detector in vacuum. Therefore
while our virial temperature is a gravitational one ascribed to the spacetime 
by the stress-energy tensor, the one of Davies-Unruh is not, it cannot be
derived from the Einstein field equations since the detector is not following 
a geodetic of the spacetime.

The fluids in nature (photon liquid, electron liquid, neutron liquid, ...) carry a 
temperature which through the stress-energy tensor determines the 
``virial temperature'' of spacetime \cite{Fantoni24f} which in turn 
excites the pure state of the quantum vacuum stimulating particle-antiparticle 
production and recombination. We can then talk of the temperature of the 
quantum vacuum of spacetime. 

We will now first discuss about the structural properties of the quantum vacuum 
for a real massless scalar field permeating the spacetime of a Local Lorentz Frame 
(LLF) and later extend our discussion to the case of General Relativity (GR).
From our recent work \cite{Fantoni24f} follows that 
$\langle R\rangle_\beta\approx 16\pi G\calt/c^4\bar{\upsilon}$ where the term on the 
left hand side is the thermal average of $R$, the Ricci curvature scalar, and, 
on the right hand side, $G$ is Newton universal gravitation constant, 
$c$ is the speed of light, $\calt$ is our ``virial temperature'' of spacetime and 
$\bar{\upsilon}$ is a constant carrying the dimensions of length squared divided 
by energy. So in flat space $R=0$, $\calt=0$ and the Minkowsky spacetime quantum 
vacuum structure will not depend on temperature. On the other hand in GR we will
be able to determine an approximation that takes care of the dependence of the 
vacuum spacetime structure from our temperature $\calt$. 

%%%%%%%%%%%%%%%%%%%%%%%%%%%%%%%%%%%%%%%%%%%%%%%%%%%%%%%%%%%%%%%%%%%%%%%%%%%%%%
\section{Discussion}
%%%%%%%%%%%%%%%%%%%%%%%%%%%%%%%%%%%%%%%%%%%%%%%%%%%%%%%%%%%%%%%%%%%%%%%%%%%%%%
\label{sec:discussion}

In Ref. \cite{Wang2017} the simple case of a real massless scalar field in flat 
Minkowski spacetime is considered
\footnote{Note that the integral measure here is chosen to be not Lorentz 
invariant in order to simplify the later structure calculation of the vacuum.}
\bq \label{eq:field}
\phi(t,\rr)=\int\frac{d\kk}{(2\pi)^{3/2}\sqrt{2\omega}}
\left[a_\kk e^{-i(\omega t-\kk\cdot\rr)}+a^\dagger_\kk e^{i(\omega t-\kk\cdot\rr)}\right],
\eq
where the temporal frequency $\omega$ and the spatial frequency $\kk$ are 
related by $\omega=|\kk|$ in natural units $\hbar=c=1$.  

The vacuum pure state $|0\rangle$ is defined by
\bq
a_\kk|0\rangle = 0~~~\forall \kk
\eq
and $a^\dagger_\kk|0\rangle = |\omega,\kk\rangle$ with 
$\langle\omega,\kk'|\omega,\kk\rangle=\delta(\kk-\kk')$ so that the
vacuum expectation value of the square modulus of the field,
$\langle\phi^2(t,\rr)\rangle_0=\langle 0|\phi^2(t,\rr)|0\rangle=\Lambda^2/8\pi^2$
with $|\kk|=\Lambda$ a high-energy (ultraviolet) cutoff. In Eq. 
(\ref{eq:field}) the first term creates an antiparticle in the sense of Dirac and the 
second a particle.

The vacuum state is an eigenstate of the Hamiltonian 
${\cal H}=\int d\rr\,T_{00}=\frac{1}{2}\int d\kk\omega\,(a_\kk a^\dagger_\kk+a^\dagger_\kk a_\kk)$,
where $T_{00}(t,\rr)=\frac{1}{2}[\dot{\phi}^2+(\nnabla\phi)^2]$. But it is 
{\sl not} an eigenstate of the energy density $T_{00}$. This fact gives rise 
to a non trivial vacuum structure. Direct calculation (See appendix A of Ref.
\cite{Wang2017}) shows that the pair correlation function
\bq \label{eq:pcf}
g(x,x')=1-\frac{\rho^{(2)}(x,x')}{\frac{2}{3}\rho^2},
\eq
where $x=(x^0,x^1,x^2,x^3)=(t,\rr)$, $x'=(t',\rr')$ are two spacetime events and
\begin{subequations}
\begin{align}
\label{eq:rho}
\rho=\rho^{(1)}(x)&=\langle T_{00}(x)\rangle_0,\\ \label{eq:cov}
{\rm covariance}(\rho)=\rho^{(2)}(x,x')&=\langle\{[T_{00}(x)-\rho^{(1)}(x)][T_{00}(x')-\rho^{(1)}(x)]\}\rangle_0.
\end{align}
\end{subequations}
where $\{AB\}=\frac{1}{2}(AB+BA)$ for any two operators $A$ and $B$, and a simple
calculation shows that $\rho=\Lambda^4/16\pi^2$ is a constant over spacetime and can 
be considered as the energy ``density'' of spacetime vacuum. After a lenghty 
calculation we \footnote{We found a sign error in their Eq. (A3).} find the
following result
\bq \label{eq:rho2llf}
\rho^{(2)}(x,x')=\frac{1}{2}\int \frac{d\kk\,d\kk'}{(2\pi)^6}
\frac{(\omega\omega'-\kk\cdot\kk')^2}{2\omega2\omega'}
\cos[(\omega-\omega')\Delta t-(\kk+\kk')\cdot\Delta\rr],
\eq
where $\Delta t=t-t'$ and $\Delta\rr=\rr-\rr'$.
 
\begin{figure}[htbp]
\begin{center}
\includegraphics[width=8cm]{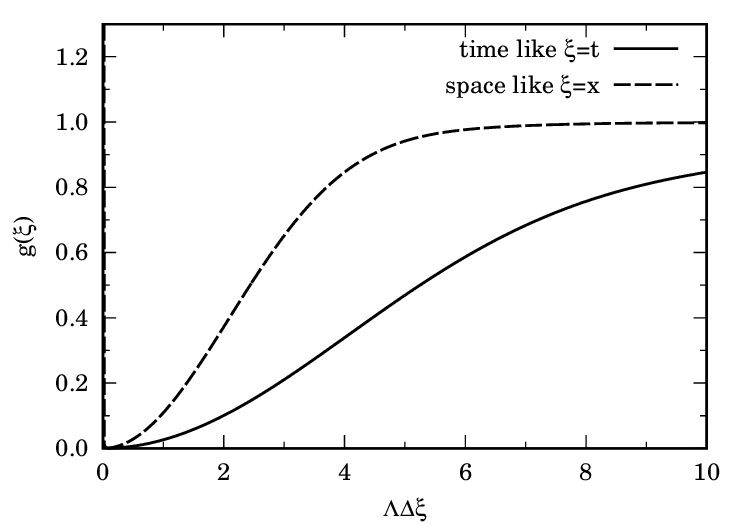}
\end{center}  
\caption{The LLF pair correlation function $g(\xi)$ of Eq. (\ref{eq:pcf}): 
when the separation of the events $x$ and $x'$ is time-like 
for $\rr=\rr'$, $\Delta\xi=|t-t'|$ and when it is space-like for $t=t'$, 
$\Delta\xi=|\rr-\rr'|$. } 
\label{fig:vs}
\end{figure}

As can be seen from Fig. \ref{fig:vs} the pair correlation function of 
(\ref{eq:pcf}) reveals an inhomogeneous and anisotropic spacetime vacuum.
It can also be easily shown that $\rho^{(2)}(x,x)=\frac{2}{3}\rho^2$ so that 
$g(0)=0$ which can be pictured as a spacetime vacuum {\sl hole} at events contact 
and on the other hand $g\to 1$ at large events separation which can be interpreted 
as a decorrelation among spacetime events of the vacuum which becomes 
{\sl uniform} and {\sl isotropic} on a large spacetime scale. From the
figure we see how both the time like pair correlation function at $|\rr-\rr'|=0$ 
and the space like one for $t-t'=0$ grow monotonously towards the uniform and 
isotropic spacetime at large events separation.  

In this picture there is no space left for a ``temperature'' of the vacuum.
Instead we expect the structure of the spacetime vacuum to feel and depend on 
temperature too.

In a recent work \cite{Fantoni24f} we introduced and defined a ``virial
temperature'' of General Relativity (GR) spacetime. Aim of the present work 
is to determine how that temperature can affect the structure of the 
quantum vacuum of spacetime.

Note that the result of Eq. (\ref{eq:pcf}) and Fig. \ref{fig:vs} looses any value 
in GR. In fact the covariance of Eq. (\ref{eq:cov}) is inherently non local and 
its calculation in a LLF will not be useful in GR. 

Now in GR the stress-energy tensor for the massless scalar field $\phi$
becomes, for a generic spacetime metric $g_{\mu\nu}$,
\bq
T_{\mu\nu}=\phi_{,\mu}\phi_{,\nu}^\dagger-\frac{1}{2}g_{\mu\nu}\phi^{,\alpha}\phi_{,\alpha}^\dagger
\eq
where a comma stands for a partial derivative and we allow the field to be 
complex for the sake of more generality.

According to Einstein field equations the stress-energy tensor of the
scalar field will induce a curvature of the spacetime
\bq \label{eq:Efe}
\langle G_{\mu\nu}\rangle_0=8\pi\langle T_{\mu\nu}\rangle_0,
\eq
where $G_{\mu\nu}$ is Einstein tensor and we are using Planck natural units
$\hbar=c=G=k_B=1$. Once again $\langle\ldots\rangle_0$ stands for a quantum
vacuum expectation value $\langle 0|\ldots|0\rangle$. In our previous work 
\cite{Fantoni24f} we defined the most natural thermal average for spacetime
that we will here indicate with the notation $\langle\ldots\rangle_\beta$
where $\beta=1/\calt$ is the inverse {\sl temperature}
\footnote{In Planck units the spacetime temperature $\calt$ varies on Planck 
energy scale $\sqrt{\hbar c^5/G}=1.9561\times 10^9$ J.}
. We will then more correctly need to average Eq. (\ref{eq:Efe}) like so
\bq
\langle\langle G_{\mu\nu}\rangle_0\rangle_\beta=8\pi\langle\langle T_{\mu\nu}\rangle_0\rangle_\beta.
\eq
Note that while the thermal average $\langle\ldots\rangle_\beta$ acts only on 
the spacetime metric $g_{\mu\nu}$ the vacuum expectation value 
$\langle\ldots\rangle_0$ acts only on the field on the right hand side of 
Eq. (\ref{eq:Efe}). On the left it has no effect and we can then rewrite
\bq
\langle G_{\mu\nu}\rangle_\beta=8\pi\langle\langle T_{\mu\nu}\rangle_0\rangle_\beta.
\eq
For example for the energy density we will find
\bq \label{eq:GRT00}
\langle\langle T_{00}\rangle_0\rangle_\beta=\langle|\dot{\phi}|^2\rangle_0-\frac{1}{2}\langle g_{00}\,g_{\mu\nu}\rangle_\beta\langle\phi_,^\mu\phi_,^{\dagger\nu}\rangle_0,
\eq
where as usual there is a hidden summation over repeated lower and upper 
indexes. 

In Ref. \cite{Fantoni24f} we were also able to render explicit the temperature.
One simply has to trace out the stress-energy tensor like so
\bq
\calt=-\frac{\bar{\upsilon}}{4}\langle T_\mu^\mu\rangle_\beta,
\eq
where
 $T_\mu^\mu=-g_{\mu\nu}\phi_,^\mu\phi_,^{\dagger\nu}$ is the stress-energy
tensor trace. Taking a vacuum expectation value of this expression we find
\bq \label{eq:temperature}
\calt=\frac{\bar{\upsilon}}{4}\langle g_{\mu\nu}\rangle_\beta\langle\phi_,^\mu\phi_,^{\dagger\nu}\rangle_0.
\eq

We will then redefine the first two $n-$points energy density correlation 
functions, now in GR
\begin{subequations}
\begin{align}
\label{eq:rhogr}
\rho=\rho^{(1)}(x)&=\langle\langle T_{00}(x)\rangle_0\rangle_\beta,\\ \label{eq:covgr}
{\rm covariance}(\rho)=\rho^{(2)}(x,x')&=\langle\langle\{[T_{00}(x)-\rho^{(1)}(x)][T_{00}(x')-\rho^{(1)}(x)]\}\rangle_0\rangle_\beta,
\end{align}
\end{subequations}
which extend Eqs. (\ref{eq:rho})-(\ref{eq:cov}) to full GR. We 
will assume that the field can still be written as in Eq. (\ref{eq:field}). 
We can then again easily calculate the two vacuum expectation values in 
Eq. (\ref{eq:GRT00}) 
\bq
\langle|\dot{\phi}|^2\rangle_0&=&\frac{\Lambda^4}{16\pi^2},\\
\langle\phi_,^\mu\phi_,^{\dagger\nu}\rangle_0&=&\int\frac{d\kk}{(2\pi)^3}\frac{p^\mu p^\nu}{2p^0},
\eq
where $p=(p^0,p^1,p^2,p^3)=(\omega,\kk)$ is the four momentum. But we
will now follow a different route. We will make the following 
approximation in Eq. (\ref{eq:GRT00})
\bq \label{eq:approx1}
\langle g_{00}\,g_{\mu\nu}\rangle_\beta\approx\langle g_{00}\rangle_\beta\langle g_{\mu\nu}\rangle_\beta
\eq 
which allows to use the result of Eq. (\ref{eq:temperature}) to find 
\bq
\rho=\langle|\dot{\phi}|^2\rangle_0-\frac{2\calt}{\bar{\upsilon}}\langle g_{00}\rangle_\beta.
\eq
Assuming furthermore that 
\bq \label{eq:approx2}
\frac{\langle g_{00}\rangle_\beta}{\bar{\upsilon}}\approx\kappa,
\eq 
a constant independent from temperature, we finally reach the following result
\footnote{Note that our virial temperature is a local quantity which can only
\cite{Fantoni24f} depend on space, so $\calt=\calt(\rr)$ in the most general 
case. In a cosmological model we will assume it to be uniform throughout the
whole Universe.}
\bq
\rho=\frac{\Lambda^4}{16\pi^2}-2\kappa\calt,
\eq
where the two approximations (\ref{eq:approx1}) and (\ref{eq:approx2}) follow a
mean tensor field spirit.

Repeating now the calculation carried on for the LLF we now find from Eq. 
(\ref{eq:covgr})
\bq \label{eq:rho2gr}
\rho^{(2)}(x,x')=-\int \frac{d\kk\,d\kk'}{(2\pi)^6}
\frac{\left(\omega\omega'-\frac{1}{2}\langle g_{00}\rangle_\beta\langle g_{\mu\nu}\rangle_\beta p^\mu p'^\nu\right)^2}{2\omega 2\omega'}
\cos[(\omega-\omega')\Delta t-(\kk-\kk')\cdot\Delta\rr],
\eq
which correctly reduces to (\ref{eq:rho2llf}) when 
$g_{\mu\nu}=\eta_{\mu\nu}={\rm diag}\{-1,1,1,1\}$.

\begin{figure}[htbp]
\begin{center}
\includegraphics[width=8cm]{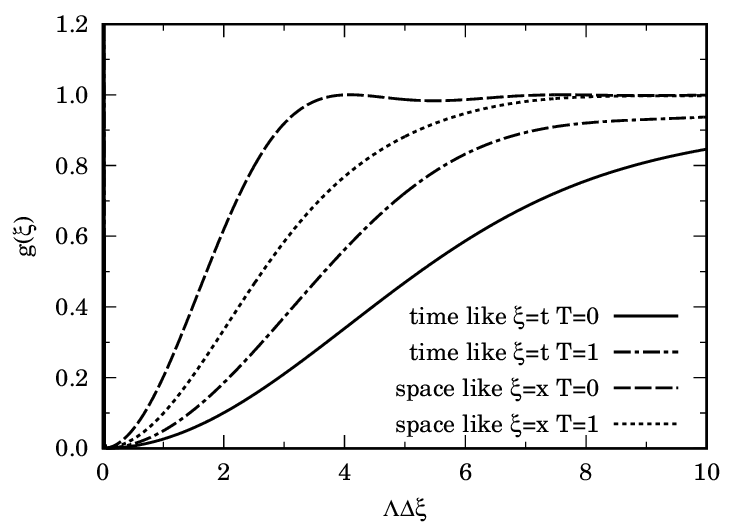}
\end{center}  
\caption{The GR pair correlation function $g(\xi)$ of Eq. (\ref{eq:pcfgr})
at two temperatures $\calt=0,1$: 
when the separation of the events $x$ and $x'$ is time-like 
for $\rr=\rr'$, $\Delta\xi=|t-t'|$ and when it is space-like for $t=t'$, 
$\Delta\xi=|\rr-\rr'|$. } 
\label{fig:vsgr}
\end{figure}

We can further think about a third approximation in order to make some progress
towards an insight on the pair correlation function of the quantum vacuum of 
full GR spacetime. a first guess could be for example the following
\bq \label{eq:rho2gr-approx}
\rho^{(2)}(x,x')\approx\int \frac{d\kk\,d\kk'}{(2\pi)^6}
\frac{\left[\omega\omega'-2\bar{\kappa}\calt\kk\cdot\kk'/(kk')\right]^2}{2\omega 2\omega'}
\cos[(\omega-\omega')\Delta t-(\kk+\kk')\cdot\Delta\rr],
\eq
where $\bar{\kappa}$ is a constant of dimension of energy. The pair
correlation function,
\bq \label{eq:pcfgr}
g(x,x')&=&1-\frac{\rho^{(2)}(x,x')}{\rho^{(2)}(x,x)},\\
\rho^{(2)}(x,x)&\approx&\left(\frac{\Lambda^4}{16\pi^2}\right)^2+\frac{\Lambda^4(\bar{\kappa}\calt)^2}{48\pi^4}.
\eq
is shown in Fig. \ref{fig:vsgr}. From the figure we see how at low temperature
the temporal structure of the spacetime quantum vacuum starts oscillating around 
the uniform and isotropic large separation limit. On the other hand the spatial
structure remains monotonic. We then see how GR allows for a ``density'' and
``temperature'' dependence of the spacetime quantum vacuum. Notwithstanding the
three approximations made and the assumption on the functional form of the 
scalar field our final result could be of some value as an application of our
Statistical Theory of Gravity \cite{Fantoni24f,Fantoni25a,Fantoni25g}. 

% 

%%%%%%%%%%%%%%%%%%%%%%%%%%%%%%%%%%%%%%%%%%%%%%%%%%%%%%%%%%%%%%%%%%%%%%%%%%%%%%
\section{Conclusions}
%%%%%%%%%%%%%%%%%%%%%%%%%%%%%%%%%%%%%%%%%%%%%%%%%%%%%%%%%%%%%%%%%%%%%%%%%%%%%%
\label{sec:conclusions}

In this brief paper we showed how the unification of the Theory of General 
Relativity and of Statistical Physics allows to treat the quantum vacuum
as a spacetime ``fluid'' with a structure which depends on the (energy)
density and on our \cite{Fantoni24f,Fantoni25a,Fantoni25g} (virial)
temperature.

This ``fluid'' could offer a clue to the search for the {\sl dark energy} 
which we today think is a missing ingredient necessary to explain 
experimental observations of our cosmos.

If we accept this picture it is natural to expect the possibility for a 
phase transition of the quantum spacetime vacuum between a gaseous, liquid, 
or crystalline phase. We can think at the spacetime of GR as the classical 
limit at high temperature when its vacuum is in a gaseous phase with 
particle-antiparticle creation and recombination close to be uniform.
In the opposite low temperature quantum limit we may expect something
like a ``superfluid'' vacuum. This should be realized by particle-antiparticle
creation and recombination filaments winding through certain regions 
of the Universe which will be inherently patchy.

\section*{Author declarations}

\subsection*{Conflicts of interest}
None declared.

\subsection*{Data availability}
The data that support the findings of this study are available from the 
corresponding author upon reasonable request.

\subsection*{Funding}
None declared.

%%%%%%%%%%%%%%%%%%%%%%%%%%%%%%%%%%%%%%%%%%%%%%%%%%%%%%%%%%%%%%%%%%%%%%%%%%%%%%
\bibliography{vf}
%\bibliographystyle{prsty}

%%%%%%%%%%%%%%%%%%%%%%%%%%%%%%%%%%%%%%%%%%%%%%%%%%%%%%%%%%%%%%%%%%%%%%%%%%%%%%
%%%%%%%%%%%%%%%%%%%%%%%%%%%%%%%%%%%%%%%%%%%%%%%%%%%%%%%%%%%%%%%%%%%%%%%%%%%%%%
%%%%%%%%%%%%%%%%%%%%%%%%%%%%%%%%%%%%%%%%%%%%%%%%%%%%%%%%%%%%%%%%%%%%%%%%%%%%%%
\end{document}